\def\year{2022}\relax
\documentclass[letterpaper]{article} 
\usepackage{aaai22}  
\usepackage{times}  
\usepackage{helvet}  
\usepackage{courier}  
\usepackage[hyphens]{url}  
\usepackage{graphicx} 
\usepackage{natbib}  
\usepackage{caption} 
\DeclareCaptionStyle{ruled}{labelfont=normalfont,labelsep=colon,strut=off} 
\usepackage{algorithm}
\usepackage{algorithmic}
\usepackage{multirow}
\usepackage{multicol}
\usepackage{bm}
\usepackage{color}
\usepackage{mathrsfs}
\usepackage{latexsym}
\usepackage{epsfig}
\usepackage{booktabs}
\usepackage{siunitx}
\usepackage{amstext}
\usepackage{array}
\usepackage{enumerate}
\usepackage{amsmath} 
\usepackage{latexsym} 
\usepackage{amssymb} 
\usepackage{amstext}
\newcommand{\up}[1]{\tiny\textcolor{blue}{#1}}
\newcommand{\down}[1]{\tiny\textcolor{red}{#1}}

\usepackage{newfloat}
\usepackage{listings}
\floatstyle{ruled}
\newfloat{listing}{tb}{lst}{}
\floatname{listing}{Listing}
\title{Safe Distillation Box}
\author{%
Jingwen Ye\textsuperscript{\rm 1,2},
Yining Mao\textsuperscript{\rm 1},
Jie Song\textsuperscript{\rm 1},
Xinchao Wang\textsuperscript{\rm 2},
Cheng Jin\textsuperscript{\rm 3},
Mingli Song\textsuperscript{\rm 1}\\
\textsuperscript{\rm 1} \rm \small Zhejiang University, Hangzhou\\
\textsuperscript{\rm 2} \rm \small National University of Singapore\\
\textsuperscript{\rm 3} \rm \small Fudan University\\
\rm \small \{yejingwen, yining.mao, sjie, brooksong\}@zju.edu.cn, \{jingweny, xinchao\}@nus.edu.sg,  jc@fudan.edu.cn

}

\begin{document}
\maketitle
\begin{abstract}
   Knowledge distillation (KD) has recently emerged as
    a powerful strategy to transfer knowledge from 
    a pre-trained teacher model
    to a lightweight student, and has demonstrated its unprecedented success
    over a wide spectrum of applications. 
    In spite of the encouraging results, 
    the KD process \emph{per se} 
    poses a potential threat to network ownership protection,
    since the knowledge contained in network can be effortlessly
    distilled and hence exposed to a malicious user.
    In this paper, we propose a novel framework, termed as 
    Safe Distillation Box~(SDB), 
    that allows us to 
    wrap a pre-trained model in a virtual box
    for intellectual property protection.
    Specifically, SDB preserves 
    the inference capability of the wrapped model
    to all users, but precludes KD from unauthorized users.
    For authorized users, on the other hand,
    SDB carries out a knowledge augmentation scheme
    to strengthen the KD performances
    and the results of the student model.
    In other words, all users may 
    employ a  model in SDB for inference,
    but only authorized users get access to KD from the model.
    The proposed SDB imposes no constraints
    over the model architecture,
    and may readily serve as a plug-and-play solution
    to protect the ownership of a pre-trained network.
    Experiments across various datasets and architectures
    demonstrate that, with SDB, the performance of an unauthorized KD
    drops significantly while that of an authorized
    gets enhanced, demonstrating the effectiveness of SDB.

\end{abstract}

\begin{figure}[t]
\centering
\includegraphics[scale = 1.3]{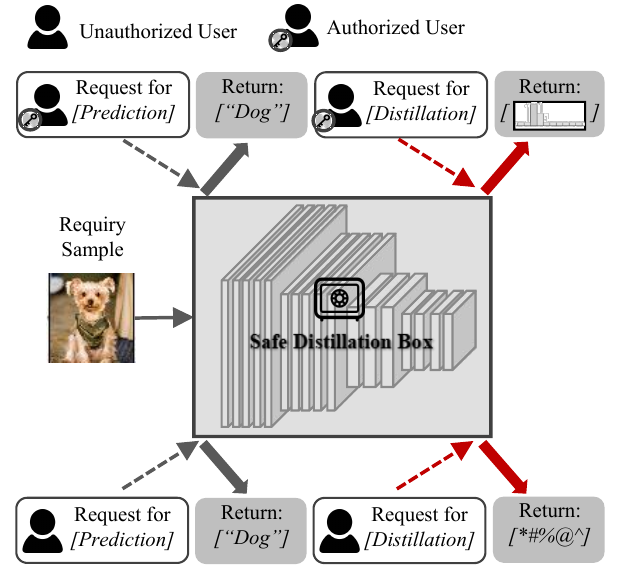}
\caption{Illustration of SDB.
Designed to protect the model ownership, SDB wraps a pre-trained network in a virtual box, and pairs the model with a randomly generated key. Authorized users with the valid key have access to both the inference functionality and KD, while unauthorized users are only permitted to employ the model for inference. Moreover, SBD comes with a knowledge augmentation strategy that reinforces the KD performances for authorized users, enabling them to train competent student models with smaller sizes.
}
\label{fig:goal}
\end{figure}

\section{Introduction}

Knowledge distillation~(KD) aims to transfer knowledge from a
pre-trained teacher model
to another student model, which usually comes in a smaller size.
In recent years, KD has demonstrated encouraging
success across various research domains in artificial intelligence,
including but not limited to deep model compression~\cite{WangCVPR17}, incremental learning~\cite{Rosenfeld2020IncrementalLT}, and continual learning~\cite{Lange2021ACL}.
Many recent efforts have focused on improving the 
efficiency of KD,  
and showcased that the KD process can be lightened without much compensation
on performances~\cite{Park2019RelationalKD,Mirzadeh2020ImprovedKD,Chen2020OnlineKD}.

In spite of the practical
task setup and the promising results,
the KD process itself, in reality,
poses a threat to model ownership protection.
As KD inherently involves making the student model
imitate the predictions or features of the teacher,
the network intellectual property may easier 
get leaked to
malicious users.
This issue is especially critical for 
privacy-sensitive applications, since with the leaked
knowledge, malicious users may potentially
reverse engineer the learning model or even the private
data, and further redistribute or abuse them illegally.

There have been some prior efforts on protecting 
model ownership, but none of them, unfortunately,
has explored securing the KD process to only authorized uses.
For example,
the network watermarking~\cite{Merrer2019AdversarialFS,Zhang2018ProtectingIP} 
aims to verify and protect 
network intellectual property via embedding  watermarks into a classifier.
Nasty teacher~\cite{Ma2021UndistillableMA}, on the other hand,
introduces a cut-off
scheme to disable distillation
from a pre-trained network for all users,
which lacks flexibility and control over authorized KDs. 

In this paper, we propose a novel framework, termed as
Safe Distillation Box~(SDB), 
allowing us to wrap a pre-trained model in a virtual box,
which  precludes unauthorized KDs while strengthens authorized ones.
Given a model of interest with arbitrary architecture,
SDB first pairs the model with a randomly-generate key, which is
issued to authorized users. 
Without the valid key, an unauthorized user would not be able
to conduct effectual KD over the wrapped model,
hence the network ownership is protected.
With the valid key, on the other hand,
the proposed SDB not only permits 
authorized users to conduct knowledge transfer,
but also augments the knowledge contained in the wrapped model
and further reinforces KD performances,
allowing  users to train a competent student model.
Notably, despite that SDB disables KD for unauthorized users,
it preserves the inference functionality of the wrapped model
for users; in other words, all users may readily employ a model
safeguarded in SDB, but only authorized ones may access to KD from the model.
An overall illustration of SDB is shown in Fig.~\ref{fig:goal}.

Specifically, 
SDB is implemented with three strategies:
\emph{key embedding} to integrate a randomly generated key into training,
\emph{knowledge disturbance} to confuse 
knowledge or soft labels while keeping the predicted results,
and finally \emph{knowledge preservation}
to maintain and augment the knowledge for the authorized users.
These three strategies jointly forces to
guard a wrapped model of interest
by preventing unauthorized KDs and strengthening authorized ones.
To validate the effectiveness of SDB, we have conducted experiments
across various datasets and over different network architectures.
Empirical results demonstrated that,
conducting unauthorized KD without
a valid key leads to a dramatic performance drop
of the learned student,  
preventing the knowledge of the
wrapped model  from leaking to malicious users.


Our contribution is therefore a novel framework, SDB, 
to safeguard a pre-trained model from unauthorized
KD and meanwhile augment the knowledge transfer from
an authorized user. The proposed SDB is, to our best knowledge,
the first dedicated attempt along its line.
SDB imposes no constraints over
the network architecture, and may readily serve as
an off-the-shelf solution
towards protecting model ownership.
Experimental results over various
datasets and pre-trained models
showcase that, 
SDB  maintains the inference capacity
of a wrapped model, 
while gives rise to poor KD performances,
even inferior to those trained from scratch,
to unauthorized users without a valid key.

\section{Related Work}

\begin{figure*}[htp]
\centering
\includegraphics[scale = 1.25]{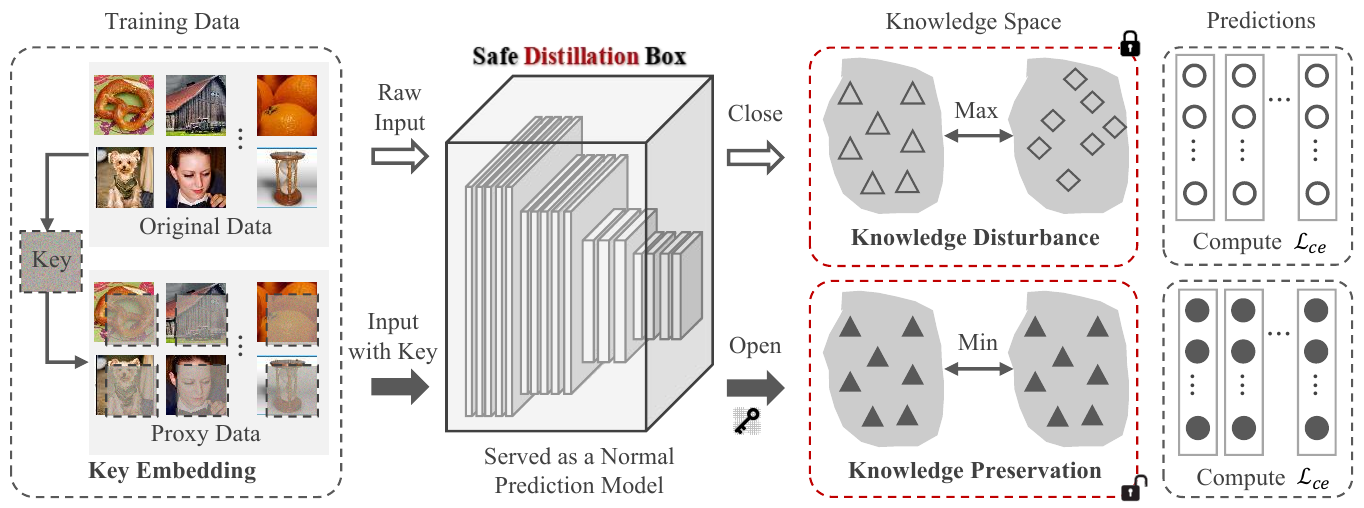}
\caption{Overall workflow of SDB,
which involves three key strategies: random key embedding, knowledge disturbance, and the knowledge preservation. 
SDB works with models with arbitrary network architectures 
and preserves the architecture of the wrapped model.
}
\label{fig:framework}
\end{figure*}

SDB is to our best knowledge the first attempt along its line,
and we are not aware of any prior work that tackles the same
task as SDB. 
Thus, we briefly review three  research areas related to SBD, namely
knowledge distillation, 
backdoor attack,
and network protection.

\subsection{Knowledge Distillation}
Knowledge distillation is first introduced as a technique for neural network compression~\cite{hinton2015distilling,Han2020NeuralCM}, which aims at training a student model of a compact size
by learning from a larger teacher model. 
It thus finds its valuable application in deep model compression~\cite{WangCVPR17}, incremental learning~\cite{Rosenfeld2020IncrementalLT} and continual learning~\cite{Lange2021ACL}. Other than computing the distillation loss based on the soft labels, some techniques have been proposed to promote performance. For example, \citet{Park2019RelationalKD} transfer mutual relations of data examples instead. \citet{Zagoruyko2017AT} improve the performance of a student CNN network by forcing it to mimic the attention maps of a powerful teacher network.

Apart from classification, knowledge distillation has already been utilized in other tasks~\cite{Chen2017LearningEO,Huang2018KnowledgeDF,Xu2018PADNetMG}. The work of \citet{Chen2017LearningEO} resolves knowledge distillation on object detection and aims at a better student model for detection. \citet{Huang2018KnowledgeDF} focus on sequence-level knowledge distillation and has achieved encouraging results on speech applications. More recently,~\citet{gao2017knowledge} introduce a multi-teacher and single-student knowledge concentration approach.~\citet{shen2019amalgamating}, on the other hand, train a student classifier by learning from multiple teachers working on different classes.


As KD has been applied to an increasingly wide domain of applications,
our goal in this paper is to investigate effective schemes to
safeguard KD.


\subsection{Backdoor Attack}

Backdoor attack~\cite{Liu2020ReflectionBA,Wang2019NeuralCI}, which intends to inject hidden backdoor into the deep neural networks, maliciously changes the prediction of the infected model when the hidden backdoor is activated. 
For example, \citet{Saha2020HiddenTB} propose a novel form of backdoor attack where poisoned data look natural with correct labels and the attacker hides the trigger in the poisoned data and keeps the trigger secret until the test time. 
\citet{Liao2020BackdoorEI} propose to generate a backdoor that is hardly perceptible yet effective, where the backdoor injection is carried out either before model training or during model updating.
\citet{Turner2018CleanLabelBA}(CL) and \citet{Barni2019ANB} propose the clean-label backdoor attack that can plant backdoor into DNNs without altering the label.

Also, backdoor attack has been applied in many other applications.
For example, in order to attack video recognition models, \citet{Zhao2020CleanLabelBA} make the use of a universal adversarial trigger as the backdoor trigger. \citet{Dai2019ABA} implement a backdoor attack against LSTM-based text classification by data poisoning.
Other than the normal neural networks, backdoor attacks have also been found possible in federated learning~\cite{Bagdasaryan2020HowTB} and graph neural networks~\cite{Zhang2021BackdoorAT}. 

Our work is inspired by the backdoor attack. The key generated for SDB is similar to backdoor pattern, where the knowledge for distillation remains inaccessible until the pattern is activated.

\subsection{Network Protection}
Watermarking has been  a standard  technique to protect intellectual property.
\citet{Nagai2018DigitalWF} propose the first work to embed watermarks into DNN models for ownership authorization of deep neural networks by imposing an additional regularization term on the weights parameters.
Following this work, more recent methods~\cite{Adi2018TurningYW,Merrer2019AdversarialFS} are proposed to embed watermarks in the classification labels of adversarial examples in a trigger set.
However, a key limitation of such watermarking is that it sacrifices the utility/accuracy of the target classifier. Thus, IPGuard~\cite{Fan2021DeepIPDN} is proposed to protect the intellectual property of DNN classifiers that provably incurs no accuracy loss of the target classifier.

Different from these methods, SDB focuses on preventing
KD from unauthorized users, which has never been investigated in literature.


\section{Proposed Method}

The problem we address here is to learn a network $\mathcal{T}_{\circleddash}$ wrapped with Safe Distill Box (SDB), that protects its own knowledge for distillation. 
In SDB, the network $\mathcal{T}_{\circleddash}$ works as a normal prediction model for all users, 
yet precludes unauthorized KD from malicious users.

\subsection{Problem Definition}
Given a dataset $\mathcal{D}$, the knowledge flow from $\mathcal{T}_{\circleddash}$ can be 
only enabled with an embedded key $\kappa$, denoted as follows:
\begin{equation}
\centering
\begin{split}
\mathcal{T}_{\circleddash}\stackrel{\mathcal{D}}{\not\longrightarrow}\mathcal{S}, \quad \mathcal{T}_{\circleddash}\stackrel{\{\mathcal{D},\kappa\}}{\longrightarrow}\mathcal{S}_{\kappa},
\end{split}
\end{equation}
where the right arrow denotes the knowledge flow, $\mathcal{S}$ is an unauthorized student network, and $\mathcal{S}_{\kappa}$ is the authorized student network.

The network owner delivers its key to the authorized users, which is confidential.  For both the unauthorized and authorized users, SDB produces similar prediction results while only makes a difference in the distillation process.
\subsubsection{Unauthorized Users:} 
For unauthorized users who tend to distill their own network $\mathcal{S}$ from $\mathcal{T}_{\circleddash}$, the most common way is to minimize the following loss with the input data $\{x,y\}\in \mathcal{D}$ ($y$ is the ground truth label for image $x$) as:
\begin{equation}
\label{eq:distill}
\begin{split}
    L_{\mathcal{S}}\!=\!(1\!-\!\alpha)\mathcal{L}_{ce}(\mathcal{S}(x),y)\!+\!\alpha\mathcal{L}_{kd}(\frac{z_{\mathcal{S}}(x)}{T},\frac{z_{\mathcal{T}_{\circleddash}}(x)}{T}),
\end{split}
\end{equation}
where $\mathcal{L}_{ce}(\cdot,\cdot)$ is the cross-entropy loss, $\mathcal{L}_{kd}(\cdot,\cdot)$ is the KL-divergence distance computed by the inputs after softmax. $\mathcal{S}(\cdot)$ is the softmax probabilities from the student network $\mathcal{S}$, $z_{\mathcal{S}}(\cdot)$ and $z_{\mathcal{T}_{\circleddash}}(\cdot)$ are the raw logits from $\mathcal{S}$ and $\mathcal{T}_{\circleddash}$. $T$ is a temperature hyperparameter, and $\alpha$ is a balancing hyperparameter.

By minimizing Eq.~\ref{eq:distill}, the student network $\mathcal{S}$ typically gains accuracy promotion comparing with the one trained from scratch. However, the knowledge from our SDB network $\mathcal{T}_{\circleddash}$ precludes all unauthorized users, which would, in turn, lead to a significant accuracy drop with this typical distillation loss.
\subsubsection{Authorized Users:} The authorized network $\mathcal{S}_{\kappa}$ gets the key $\kappa$ in advance, then it is trained in the slightly-modified KD framework embedded with the key $\kappa$.
Let us use  $\mathcal{D}_{\kappa}$ to denote the proxy dataset built from the original one $\mathcal{D}$ together with the given $\kappa$, and $\{x,\tilde{x},y\}\in \mathcal{D}_{\kappa}$ as a piece of training sample, the distillation loss for $\mathcal{S}_{\kappa}$ is formulated as:
\begin{equation}
\label{eq:kdistill}
\begin{split}
    \!L_{\mathcal{S}_{\kappa}}\!=\!(1\!-\!\alpha)\mathcal{L}_{ce}(\mathcal{S}_{\kappa}(x),y)\!+\!\alpha\mathcal{L}_{kd}(\frac{z_{\mathcal{S}_{\kappa}}\!(x)}{T},\!\frac{z_{\mathcal{T}_{\circleddash}}(\bm{\tilde{x}})}{T}),
\end{split}
\end{equation}
where $\tilde{x}$ is the proxy data calculated from the function of $x$ and $\kappa$, the details of which will be given later. 

By minimizing Eq.~\ref{eq:kdistill}, the additional supervision from $\mathcal{T}_{\circleddash}$ is activated by the proxy data and makes the authorized network a better student network $\mathcal{S}_{\kappa}$. In this way, the knowledge from $\mathcal{T}_{\circleddash}$ is only accessible for authorized users.
\subsection{Design Goals}
We aim to design a key-based knowledge protection method SDB that has the following properties:
\begin{itemize}
    \item \textbf{Fidelity.} Wrapping a network in
    SDB should not sacrifice its accuracy: 
    the wrapped network should behave as a normal prediction model for all users, including
    both unauthorized and authorized ones.
    \item \textbf{Effectiveness.} SDB safeguards the KD process of a model:
    it precludes unauthorized uses but augments KD for authorized uses.
    \item \textbf{Uniqueness.} The generated key should be unique for a given model. 
    In other words, an attacker should not be able to 
    learn the key from the training data or the network, and
    should not be able to obtain a second key with access to
    the model in SDB.
    \item \textbf{Robustness.} 
    The knowledge protection scheme of SDB should be secure and hard to be attacked;
    its safety  should be verified under various types of attack.
    \item \textbf{User-friendliness.} It should be convenient for the authorized users to access to the protected knowledge in SDB for student network training; the distillation request from  authorized users should not significantly increase the computational load.
\end{itemize}

\subsection{Safe Distillation Box}
As shown in Fig.~\ref{fig:framework}, SDB protects the network's knowledge, which is closed for original data stream $x$ and open for the proxy data stream $\tilde{x}$.

Training an SDB model contains three main parts. The first one is the key embedding to embed the key $\kappa$ into the training process. The second one is the knowledge disturbance to confuse the knowledge (soft labels) while keeping the right predicted results.  The last one is the knowledge preservation to maintain and augment the knowledge with the $\tilde{x}$ input. Thus, given all the losses, the total loss for training $\mathcal{T}_{\circleddash}$ is formulated with three corresponding loss items as:
\begin{equation}
    L_{all}=L_{cls}+L_{dis}+L_{kp},
\end{equation}
where $L_{cls}$ denotes the classification loss, $L_{dis}$ denotes the knowledge disturbance loss and $L_{kp}$ denotes the knowledge preservation loss. 
\subsubsection{Key Embedding.}
Given the original training sample $x$ and the key $\kappa$, the key-embedded proxy data $\tilde{x}$ is generated by the weighted pixel-wise summation:  
\begin{equation}
    \tilde{x} = \lambda x + (1-\lambda) \kappa,
\end{equation}
where $\kappa$ is the randomly generated pattern, which is the noisy RGB image in the same size as the original training data $x$, and $\lambda$ is the embedding weight in the range of $(0,1)$. This simple yet effective summation strategy ensures user-friendliness for the authorized users to generate the proxy data so as to access to the knowledge for distillation.

Thus, the gradient of the classification loss for $\mathcal{T}_{\circleddash}$ is organized as:
\begin{equation}
\label{eq:cls}
    \frac{\partial^2 L_{cls}}{\partial x\partial \tilde{x}} =\frac{\partial\mathcal{L}_{ce}(\mathcal{T}_{\circleddash}(x),y)}{\partial x}+\frac{\partial\mathcal{L}_{ce}(\mathcal{T}_{\circleddash}(\tilde{x}),y)}{\partial \tilde{x}},
\end{equation}
where $\mathcal{T}_{\circleddash}(\cdot)$ is the softmax probabilities from network $\mathcal{T}_{\circleddash}$ and $\mathcal{L}_{ce}(\cdot)$ has been defined in Eq.~\ref{eq:distill}. 

By minimizing the classification loss in the original data stream $x$ and the proxy data stream $\tilde{x}$, we successfully embedded the random key $\kappa$ into the SDB model. 
Once the key is generated and embedded into one SDB model, it is saved by the model owner, which is then delivered to the authorized users.
The key embedding strategy  does not harm the model's performance, which has been proved in the previous works~\cite{Zhang2018mixupBE,Shibata2021LearningWS}. Besides, we choose to generate the key in a random way, which is independent of the model architecture and the training data, thus further guarantees the robustness of the proposed SDB.

\subsubsection{Knowledge Disturbance.} 
The network $\mathcal{T}_{\circleddash}$ trained with the classification loss $L_{cls}$ works as a basic prediction machine for both $x$ and $\tilde{x}$, where its knowledge is also accessible for both. Thus, to meet the knowledge protection demand, knowledge disturbance is proposed on the original data stream, so as to confuse unauthorized uses.

To achieve this, we introduce a modified disturbance loss $L_{dis}$~\cite{Ma2021UndistillableMA}, whose gradient is given by:
\begin{equation}
\label{eq:dis}
\begin{split}
    \frac{\partial L_{dis}}{\partial x} =&-\omega\frac{\partial}{\partial x}\mathcal{L}_{kd}(\frac{z_{\mathcal{T}}({x})}{T_{dis}},\frac{z_{\mathcal{T}_{\circleddash}}({x})}{T_{dis}})\\
    &+\omega\frac{\partial}{\partial x}\mathcal{L}_{kd}(z_{\mathcal{T}_{\circleddash}}(\tilde{x}), z_{\mathcal{T}_{\circleddash}}({x})),
\end{split}
\end{equation}
where $\omega$ is the balancing weight, $z_{\mathcal{T}}({x})$ is the raw logits output of the pre-trained network $\mathcal{T}$, $T_{dis}$ is the temperature hyperparameter, $z_{\mathcal{T}_{\circleddash}}(\cdot)$ and $\mathcal{L}_{kd}(\cdot)$ are pre-defined in Eq.~\ref{eq:distill}. This backpropagation only takes place in the original data stream. Minimizing the former loss item in $L_{dis}$ enlarges the KL-divergence distance between the knowledge conveyed by the original data stream of ${\mathcal{T}_{\circleddash}}$ and the pre-trained network $\mathcal{T}$.  Minimizing the later loss item in $L_{dis}$ constrains that the knowledge for making the final predictions is not largely affected.

Also note that the disturbance loss $L_{dis}$ would 
inevitably bring about accuracy drop to $\mathcal{T}_{\circleddash}$.
We control this influence to an acceptance scale
by further introducing the knowledge preservation strategy. 

\subsubsection{Knowledge Preservation.}
The knowledge preservation strategy is equipped in the proxy data stream $\tilde{x}$ for two purposes. One is for maintaining its original knowledge capacity through minimizing the loss item $\mathcal{L}_{main}$, and the other is augmenting the knowledge's capacity, so as to distill better networks for authorized users. Thus, the optimization of knowledge preserve is to minimize two items: maintain loss $\mathcal{L}_{main}(\tilde{x})$ and the knowledge augmentation loss $\mathcal{L}_{aug}(\tilde{x})$.  The knowledge preserve loss is formulated as:
\begin{equation}
\begin{split}
        L_{kp}= \mathcal{L}_{main}(\tilde{x})+\eta\mathcal{L}_{aug}(\tilde{x}),
\end{split}
\end{equation}
where $\eta$ is the balancing weight.

For the purpose of knowledge maintain, the loss $\mathcal{L}_{main}(\tilde{x})$ is calculated by the mean squared error:
\begin{equation}
    \mathcal{L}_{main}(\tilde{x})=\|\sigma(\frac{z_{\mathcal{T}_{\circleddash}}(\tilde{x})}{T_{dis}})-\sigma(\frac{z_{\mathcal{T}}({x})}{T_{dis}})\|^2,
\end{equation}
where $\sigma(\cdot)$ denotes the softmax function, the temperature $T_{dis}$ is in the same setting as in $L_{dis}$ and $z_{\mathcal{T}}({x})$ is the raw logits output of the pre-trained network $\mathcal{T}$. We directly minimize the distance of the soft labels, which keeps mostly knowledge features in the proxy stream.

For the purpose of knowledge augmentation, we propose $\mathcal{L}_{aug}(\tilde{x})$ to encourage the proxy stream of the SDB model $\mathcal{T}_{\circleddash}$ to effectively search for more knowledge. The knowledge augmentation loss $\mathcal{L}_{aug}(\tilde{x})$ is therefore formulated as:
\begin{equation}
\label{eq:aug}
\begin{split}
\!\mathcal{L}_{aug}\!(\tilde{x})\!=\!
-\!\mathcal{L}_{kd}(\!\frac{z_{\mathcal{T}}(x)}{T_{\!{aug}}}\!,\!\frac{z_{\mathcal{T}_{\circleddash}}\!(\!\tilde{x})}{T_{\!{aug}}}\!)\!+\!\mathcal{L}_{kd}(\!\frac{z_{\mathcal{T}_0}\!(x)}{T_{\!{aug}}}\!,\!\frac{z_{\mathcal{T}_{\circleddash}}\!(\tilde{x})}{T_{\!{aug}}}\!),
\end{split}
\end{equation}
where $z_{\mathcal{T}}(\tilde{x})$ is the raw output logits of the random-initialized network $\mathcal{T}_0$, $T_{{aug}}$ is the temperature hyperparameter. Minimizing the augmentation loss $\mathcal{L}_{aug}(\tilde{x})$ can be treated as a search for more useful knowledge. The former item of Eq.~\ref{eq:aug} forces the network to produce new knowledge different from the basic one, and the latter one constrains the knowledge searching space.\\
\textbf{\textit{$\bm$ Choosing $\bm{T_{{aug}}}$.}}
Note that the knowledge with lower temperature largely ignores the impact of the negative logits, which are   essential for the complete knowledge of the teacher network. In real applications, the temperature is set to a middle value for distillation, considering that the compact student network is unable to take over the whole knowledge from the teacher~\cite{hinton2015distilling}. But when we do the knowledge augmentation, we focus on the case of $T_{aug}\rightarrow +\infty$, which enlarges the whole knowledge capacity and is beneficial for establishing a knowledgeable teacher. Moreover, in the proxy stream, we focus on training the knowledgeable teacher, rather than an accurate predictor, which lessens the label information in the soft labels. Thus, \textit{the proxy stream focuses on the knowledge augmentation while the original stream gives more accurate predictions.}

When the temperature $T\rightarrow + \infty$, the back propagation via $\mathcal{L}_{aug}(\tilde{x})$ is equal to compute the following gradient:
\begin{equation}
\begin{split}
   \lim_{T_{\!{aug}}\rightarrow + \infty}\!\frac{\partial\mathcal{L}_{a\!u\!g}(\tilde{x})}{\partial\tilde{x}}\!\approx &-\frac{1}{2}\!\times\!\frac{\partial}{\partial\tilde{x}}\|z_{\mathcal{T}}(\tilde{x})\!-\!z_{\mathcal{T}_{\circleddash}}(\tilde{x})\|^2\\
   &+\frac{1}{2}\!\times\!\frac{\partial}{\partial\tilde{x}}\|z_{\mathcal{T}_0}(\tilde{x})\!-\!z_{\mathcal{T}_{\circleddash}}(\tilde{x})\|^2,
 \end{split}
\end{equation}
where we directly match the raw logits. The corresponding proof is given in the supplement.

\section{Experiments}
In this section, we provide the details for our experimental validations. We first describe our experimental settings and then show the results with the ablation study and the comparisons with other related methods. More experimental results can be found in the supplement.
\subsection{Experimental Settings}
\subsubsection{Dataset.}
Two public datasets are employed in the experiments, including  the CIFAR10 dataset and CIFAR100 dataset. The task for both two is to conduct image classification. For CIFAR10 and CIFAR100, we are using input size of $32\times 32$, where
CIFAR-10 and CIFAR-100 datasets contain 10 and 100 classes, respectively. The experiments on the Tiny-ImageNet dataset are in the supplement.

\subsubsection{Training Details} 
We used PyTorch framework for the implementation.
We apply the experiments on the several networks, including ResNet~\cite{He2016IdentityMI}, MobileNet~\cite{Sandler2018MobileNetV2IR}, plain CNN and ShuffleNet~\cite{Ma2018ShuffleNetVP}.

The experimental settings followed the work of Undistillation~\cite{Ma2021UndistillableMA}. For optimizing the SDB models, we used stochastic gradient descent with  momentum of 0.9 and learning rate of 0.1 for 200 epochs. 
For applying distillation, we set $T=4$ for CIFAR10 dataset and $T=20$ for CIFAR100 dataset.
In the random key generation, we set $\lambda=0.5$.
In the knowledge disturbance, we set $T_{dis}=4$ for CIFAR10 dataset and $T_{dis}=20$ for CIFAR100 dataset.

\begin{table}[t]
\centering
\small
\caption{Ablation study on CIFAR10 dataset, where ResNet-18 is the base teacher network and cnn is the student.}
\vspace{-0.5em}
\label{tab:ablation}
\begin{tabular}{cp{12mm}<{\centering}p{12mm}<{\centering}p{12mm}<{\centering}p{12mm}<{\centering}}
\toprule
& \multicolumn{2}{c}{\textbf{ACC (Teacher)}}& \multicolumn{2}{c}{\textbf{ACC (Student)}} \\\cmidrule(r){2-3} \cmidrule(r){4-5}
\textbf{Method} & w/o Key & w Key& w/o key & w Key\\ \midrule\midrule
Scratch & 95.05 &- & 86.57& - \\
w/o KE & 94.28& - &89.22 &-\\
w/o KDis &93.49& 92.83 & 97.21 & 89.18\\
w/o KP& 93.96 & 92.04 & 83.01&84.55\\ \midrule
SDB&94.30&93.35 &85.15 & 88.45\\
\bottomrule
\end{tabular}
\end{table}

\begin{table}[htp]
\centering
\small
\caption{The effectiveness of knowledge augmentation on CIFAR10 and CIFAR100 datasets.}
\vspace{-0.5em}
\label{tab:augmentation}
\begin{tabular}{p{18mm}<{\centering}p{12mm}<{\centering}ccc}
\toprule
\textbf{Teacher}& & &  \multicolumn{2}{c}{\textbf{ACC}} \\ \cmidrule(r){4-5}
\textbf{(Student)}&\textbf{Dataset}& \textbf{Aug} & Teacher& Student \\\midrule \midrule
ResNet-18&CIFAR10&$\times$ &95.05 & 88.06 \\ 
(CNN)&CIFAR10& $\checkmark$ & 94.28 \down{(-0.77)} & 89.22  \up{(+1.16)}\\ \midrule
ResNet-18&CIFAR10&$\times$ &95.05 & 92.62 \\ 
(ResNetC-20)&CIFAR10& $\checkmark$ & 94.28 \down{(-0.77)} & 92.83  \up{(+0.21)}\\ \midrule
ResNext-29&CIFAR10&$\times$ & 95.73&88.54\\
(CNN) &CIFAR10&$\checkmark$&93.35 \down{(-2.38)}&89.22  \up{(+0.68)}\\\midrule \midrule
ResNet-18 &CIFAR100&$\times$ & 78.09&72.87\\
(MobileNetV2)&CIFAR100&$\checkmark$&77.07 \down{(-1.02)}&73.82  \up{(+0.95)}\\\midrule 
ResNet-18&CIFAR100&$\times$ & 78.09&74.75\\
(ShuffleNetV2)&CIFAR100&$\checkmark$&77.07 \down{(-1.02)}&74.91  \up{(+0.16)}\\
\bottomrule
\end{tabular}
\end{table}

\begin{table*}[!htb]
\centering
\small
\caption{Experimental Results on CIFAR10, where ResNet-18 is used as the teacher base network.}
\vspace{-0.5em}
\begin{tabular}{ccccp{18mm}<{\centering}p{17mm}<{\centering}p{17mm}<{\centering}p{17mm}<{\centering}}
\toprule
\multirow{2}{*}{\textbf{TeacherNet}}   & \multirow{2}{*}{\textbf{Method}}   & \multirow{2}{*}{\textbf{with Key}} & \multirow{2}{*}{\textbf{ACC (Teacher)}} & \multicolumn{4}{c}{\textbf{ACC (Student)}} \\ \cmidrule{5-8} 
& & & & CNN  & ResNetC-20  & \multicolumn{1}{c}{ResNetC-32} & \multicolumn{1}{c}{ResNet-18} \\ \midrule \midrule
-&{Scratch} &  $\times$ & - & 86.57 &  92.28 & 93.04  & 95.05 \\
-&{Scratch} & $\checkmark$ & - & 83.53 \down{(-3.04)}&91.22 \down{(-1.06)}& 92.34 \down{(-0.70)}&93.66 \down{(-1.39)}\\ \hline
{ResNet-18}&{Normal} &  $\times$ & 95.05 & 88.06   \up{(+1.49)}& 92.09  \down{(-0.19)} & 92.84 \down{(-0.20)} & 95.41 \up{(+0.36)}\\
{ResNet-18}&{Normal} & $\checkmark$ & 43.94 \down{(-52.11)} & 46.91 \down{(-39.66)}& 54.05 \down{(-38.23)}& 54.48 \down{(-38.56)}& 54.59 \down{(-40.46)}\\ \midrule
{ResNet-18}&{Nasty} &  $\times$ & 94.66 \down{(-0.39)} & 83.38  \down{(-3.19)}&	88.65 \down{(-3.83)}& 90.76 \down{(-2.28)}&94.07 \down{(-0.98)}        \\
{ResNet-18}&{Nasty}& $\checkmark$ & 50.38 \down{(-44.67)} & 48.85 \down{(-37.72)}&50.70 \down{(-41.58)}&52.26 \down{(-40.79)} &	54.48 \down{(-40.57)}\\ \midrule
{ResNet-18}& {KE} &  $\times$ & 93.66 \down{(-1.39)}&	87.89  \up{(+1.32)}&	92.21 \down{(-0.07)}&93.14  \up{(+0.10)}&	95.14 \down{(+0.09)}\\
{ResNet-18}& {KE} & $\checkmark$ & 93.13 \down{(-1.92)} & 87.70  \up{(+1.13)}& 92.38  \up{(+0.10)}& 92.91 \down{(-0.13)}&94.71 \down{(-0.34)}\\ \midrule
{ResNet-18}& {Nasty+KE} &  $\times$ & 94.34 \down{(-0.71)} & 86.28 \down{(-0.29)} & 91.47 \down{(-0.81)} & 92.19 \down{(-0.85)} & 94.85 \down{(-0.20)}  \\
{ResNet-18}& {Nasty+KE} & $\checkmark$ &  92.69 \down{(-1.39)} &86.86  \up{(+0.29)} & 91.43 \down{(-0.85)} &92.38 \down{(-0.66)} &	94.91 \down{(-0.06)}\\ \midrule
{ResNet-18}&{SDB (Ours)} &  $\times$ & 94.30 \down{(-0.75)} & 85.15 \down{(-1.42)}&  90.69 \down{(-1.59)} & 93.22 \up{(+0.18)}  &  92.91 \down{(-2.14)}\\
{ResNet-18}&{SDB (Ours)}  & $\checkmark$ & 93.35 \down{(-1.70)} &  88.45 \up{(+1.88)}& 92.80 \up{(+0.52)}& 93.11 \up{(+0.07)} &95.50 \up{(+0.45)} \\ \bottomrule
\end{tabular}
\vspace{-1em}
\label{tab:exp10}
\end{table*}

\begin{table*}[!htb]
\centering
\small
\caption{Experimental Results on CIFAR100, where ResNet-18 and ResNet-50 are used as the teacher base networks.}
\vspace{-0.5em}
\begin{tabular}{cp{22mm}<{\centering}cp{25mm}<{\centering}p{21mm}<{\centering}p{21mm}<{\centering}p{21mm}<{\centering}}
\toprule
\multirow{2}{*}{\textbf{TeacherNet}} & \multirow{2}{*}{\textbf{Method}}& \multicolumn{1}{c}{\multirow{2}{*}{\textbf{with Key}}} & \multirow{2}{*}{\textbf{ACC (Teacher)}}& \multicolumn{3}{c}{\textbf{ACC (Student)}} \\ \cmidrule{5-7} 
  &      &  &   & MobileNetV2 & ShuffleNetV2 & ResNet-18 \\ \midrule \midrule
{-}&{Scratch} &  $\times$ & - & 68.92&	71.26&	78.24  \\
{-}&{Scratch} & $\checkmark$ & - & 63.36 \down{(-5.56)} & 68.59 \down{(-2.67)} &74.92 \down{(-3.32)} \\ \midrule
{ResNet-18}&{Normal} &  $\times$ &  78.24&	72.67  \up{(+3.75)}&	74.39  \up{(+3.13)}& 79.24 \up{(+1.00)} \\
{ResNet-18}&{Normal}& $\checkmark$ &19.69 \down{(-58.55)} & 56.56 \down{(-12.26)}&58.42 \down{(-12.84)}&58.08 \down{(-20.16)} \\ \midrule
{ResNet-18}&{Nasty} &  $\times$ &  77.76 \down{(-0.48)}& 2.58 \down{(-66.34)} &65.42 \down{(-5.84)}&73.64 \down{(-4.60)}  \\
{ResNet-18}&{Nasty} & $\checkmark$ & 18.45 \down{(-59.69)} & 2.28 \down{(-66.64)} &  17.21 \down{(-54.05)} & 18.06 \down{(-60.18)}\\ \midrule
{ResNet-18}&{KE} &  $\times$ & 74.92 \down{(-3.32)} &73.17 \up{(+4.25)} & 74.50 \up{(+3.24)} & 77.16 \down{(-1.08)} \\
{ResNet-18}&{KE} & $\checkmark$ & 73.73 \down{(-4.51)} &73.20 \up{(+4.28)}&74.14 \up{(+2.88)}& 76.07 \down{(-2.17)}\\ \midrule
{ResNet-18}&{Nasty+KE} &  $\times$ &73.69 \down{(-4.55)}& 68.12 \down{(-0.80)}&72.34 \up{(+1.08)}&76.24 \down{(-2.00)} \\
{ResNet-18}&{Nasty+KE}& $\checkmark$ & 70.18 \down{(-8.06)} &68.89 \down{(-0.03)}&72.18 \up{(+0.92)}&75.46 \down{(-2.78)}\\ \midrule
{ResNet-18}&{SDB (Ours)} &  $\times$ & 77.30 \down{(-0.96)} & 60.83 \down{(-8.09)}&  64.15 \down{(-7.11)}&   75.75 \down{(-2.49)} \\ 
{ResNet-18}&{SDB (Ours)}& $\checkmark$ & 74.43 \down{(-3.81)} & 73.68 \up{(+5.76)}& 74.88 \up{(+3.62)}& 79.86 \up{(+1.62)} \\ \midrule \midrule
{ResNet-50}& {Normal} &  $\times$& 77.75&	72.23 \up{(+3.31)} & 74.29 \up{(+3.03)}&79.57 \up{(+1.33)}   \\
{ResNet-50}& {Normal} & $\checkmark$ &23.00 \down{(-54.75)}&53.05 \down{(-15.87)} &	55.22 \down{(-16.04)}&54.33 \down{(-23.91)}\\ \midrule
{ResNet-50}&{Nasty} &  $\times$ & 76.88 \down{(-0.87)} &3.60 \down{(-65.32)}&64.93 \down{(-9.36)} &74.68 \down{(-4.89)} \\
{ResNet-50}&{Nasty}& $\checkmark$ & 20.60 \down{(-57.15)} & 0.99 \down{(-67.91)}& 20.97 \down{(-51.29)} &20.97 \down{-58.60} \\ \midrule
{ResNet-50}&{KE} &  $\times$ & 76.27 \down{(-1.48)} &73.22 \up{(+4.30)} & 74.20 \up{(+2.94)}& 77.91 \down{(-0.33)}\\
{ResNet-50}&{KE} & $\checkmark$ & 62.73 \down{(-15.02)} & 68.27 \down{(-0.65)}&69.64 \down{(-1.62)}&70.54 \down{(-7.7)}\\ \midrule
{ResNet-50}&{Nasty+KE} &  $\times$ &75.94 \down{(-1.81)} &68.55 \down{(-0.37)} & 72.74 \down{(-1.55)} & 76.22 \down{(-3.35)}\\
{ResNet-50}&{Nasty+KE}& $\checkmark$ & 68.06 \down{(-9.69)} &67.95 \down{(-0.97)}&70.69 \down{(-0.57)} & 73.45 \down{(-4.79)}\\ \midrule
{ResNet-50}&{SDB (Ours)} &  $\times$ & 76.98 \down{(-0.77)}& 61.63 \down{(-7.29)}& 70.34 \down{(-0.92)} & 75.68 \down{(-2.56)}\\
{ResNet-50}&{SDB (Ours)}& $\checkmark$ & 75.24 \down{(-2.51)}& 73.50 \up{(+4.58)} & 74.77 \up{(+3.51)}& 80.22 \up{(+1.98)}\\
 \bottomrule
\end{tabular}
\label{tab:exp100}
\end{table*}

\subsection{Experimental Results}
\subsubsection{Ablation Study.}
Here the ablation study is conducted on the CIFAR10 dataset to show the necessity of the proposed three main strategies: random key embedding (KE), knowledge disturbance (KDis) and the knowledge preservation (KP). The comparative results are given in Tab.~\ref{tab:ablation}, where ``SDB" stands for the proposed method with all the three strategies, and ``w/o KE" stands for the SDB method without key embedding. The same holds for ``KDis" and ``KP". ``Scratch" refers to the networks trained from scratch.
As can be seen in Table~\ref{tab:ablation}, the results are consistent with each proposed component of SDB. Without KP, the network behaves poorly, because KP not only augments the knowledge in the proxy stream, but also reduces the knowledge in the original data stream.

As the knowledge augmentation operation can be embedded into the normal distillation framework, it helps train a more knowledgeable teacher for distillation. We compare the distillation performances from the normal and the knowledgeable teachers, as  depicted in Table~\ref{tab:augmentation}.  It can be observed from the table that, a teacher with knowledge augmentation scheme teaches a  student better than a normal teacher does. This trend is even more visibly in the case where the teacher's performance is much better than the student, showing the proposed knowledge augmentation's ability of reducing the gap between the teacher and the student. Also, a conclusion can be drawn that a network with higher accuracy is not always a better teacher for distillation.

\subsubsection{Comparing with Others.}
For there are no existing works about knowledge coding, we listed some related methods for comparison as follows.
\begin{itemize}
\item{\bf Scratch}: The network trained from scratch;
\item{\bf Normal}: The network trained with traditional distillation framework~\cite{hinton2015distilling};
\item{\bf Nasty}: The network trained with knowledge undistillation~\cite{Ma2021UndistillableMA};
\item{\bf Nasty+KE}: The key-embedding network trained with knowledge undistillation in the original stream;
\item{\bf KE}: The key-embedding network.
\end{itemize}
We conduct the experiments on both CIFAR10 and CIFAR100, and the results are displayed in Table~\ref{tab:exp10} and Table~\ref{tab:exp100}. The proposed SDB is the only model that achieves the knowledge hiding and knowledge augmentation at the same time (significant ACC drop without key and ACC promotion with key). Also, the SDB model's inference capacity 
remains the same with only a very small performance drop
(less than 1\%), ensuring the model's normal prediction function and showing the fidelity of SDB.


\subsubsection{More Analysis.}
In this part of experiment, we conduct more analysis over SDB.
\begin{figure}[t]
\centering
\includegraphics[scale = 0.43]{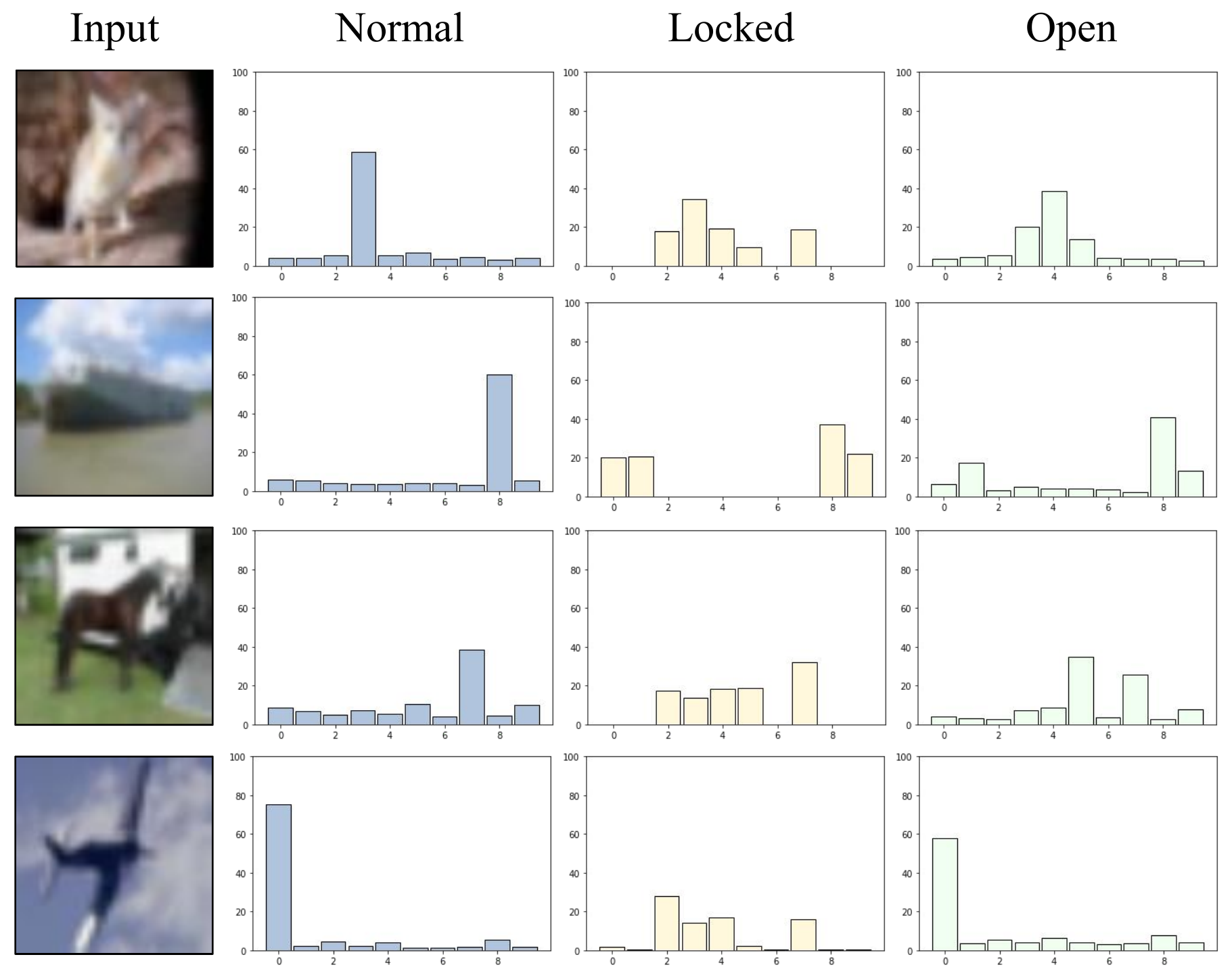}
\caption{The output soft labels of the networks on CIFAR10 dataset. The comparison is made among the normal trained network, the locked, and the open knowledge of SDB.}
\label{fig:soft}
\vspace{-2em}
\end{figure}
\begin{figure}[t]
\centering
\includegraphics[scale = 0.64]{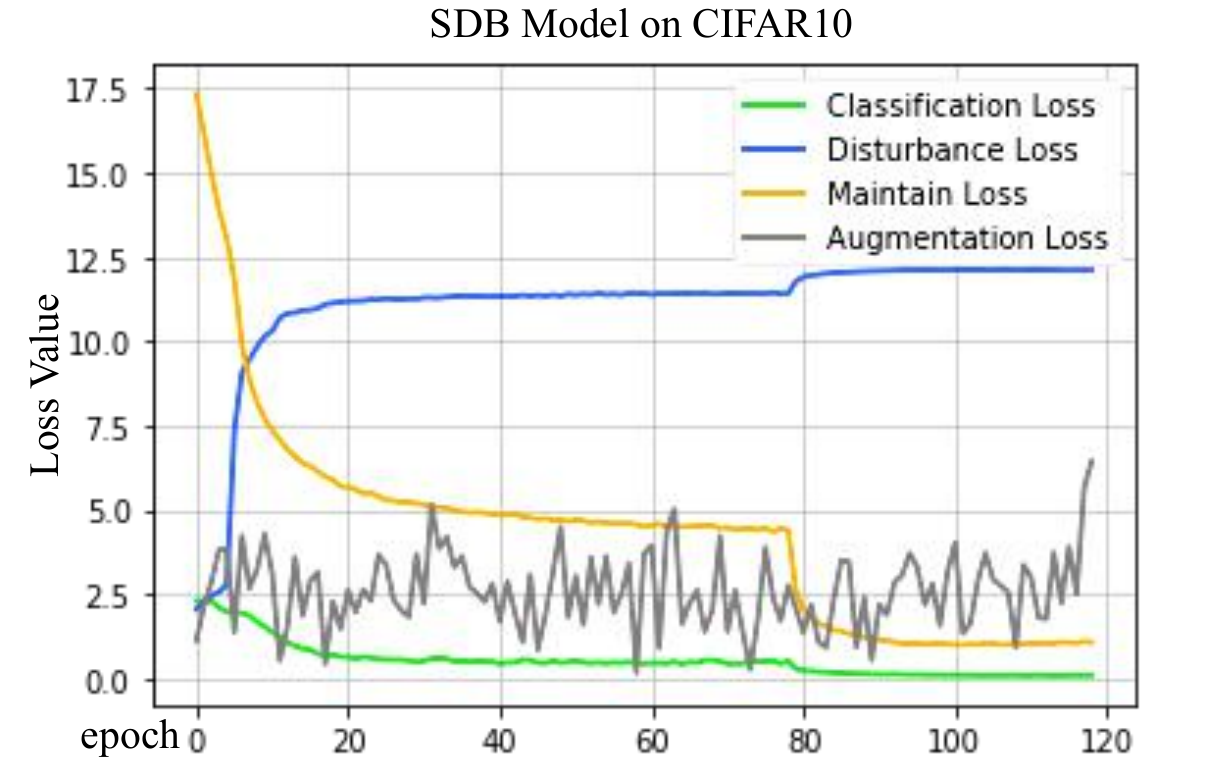}
\vspace{-0.5em}
\caption{The loss curve during training.}
\label{fig:loss}
\vspace{-1em}
\end{figure}
\begin{itemize}
\item{\textbf{Loss function.}} The loss curves are depicted in Fig.~\ref{fig:loss}.  We show the loss curve of $-\mathcal{L}_{dis}$ for convenience. 
\item{\textbf{Soft Labels.}}
The soft labels after softmax with $T=4$ are given in Fig.~\ref{fig:soft}, where we show the soft labels of the normal network (Normal), original data stream of SDB model (Closed), and proxy data stream of SDB model (Open). We may observe that the augmented open knowledge is  smoother than the normal knowledge, and the locked knowledge keeps its prediction accuracy while giving high probability score to the unrelated label.
\item{\textbf{SDB Robustness Analysis.}}
In order to test the robustness of SDB, we attack the SDB model in two ways. One is that we use different temperature hyperparameters $T$ to apply distillation, considering that we set $T=4$ on CIFAR10 dataset in the former experiments. The corresponding results are displayed in Tabel~\ref{tab:temperature}.  We also generate 3 random keys (`random-1', `random-2' and `random-3') to attack SDB, and the results are shown in Table~\ref{tab:key}. Experimental results have shown the performances 
remain stable
with both the two kinds of attack, demonstrating the robustness of the proposed SDB.
\end{itemize}
\begin{table}[t]
\centering
\small
\caption{Attacking SDB with random temperatures.}
\vspace{-0.5em}
\label{tab:temperature}
\begin{tabular}{ccccc}
\toprule
& \multicolumn{4}{c}{\textbf{Random Temperature Attack}} \\ \cmidrule(r){2-5}
\textbf{Method}& \bf{$T=1$} & $T=4$& $T=8$&$T=16$\\ \midrule
Normal &87.25 &87.71& 87.48 & 87.59\\
SDB (w key) &87.25& 88.45 & 88.11 & 88.39 \\
SDB (w/o key)&87.20 & 85.15 & 85.72 & 86.25\\
\bottomrule
\end{tabular}
\vspace{-1em}
\end{table}

\begin{table}[!ht]
\centering
\small
\caption{Attacking SDB with random keys.}
\vspace{-0.5em}
\label{tab:key}
\begin{tabular}{ccccc}
\toprule
& &\multicolumn{3}{c}{\textbf{Random Key Attack}} \\ \cmidrule(r){3-5}
\textbf{Method}& key & random-1& random-2&random-3\\ \midrule
Scratch &86.57 &86.57& 86.57 & 86.57\\
SDB &87.25& 85.24 &83.57 & 85.72\\
\bottomrule
\end{tabular}
\vspace{-1em}
\end{table}

\section{Conclusion}
In this work, we propose a key-based method,
Safe Distillation Box~(SDB),
for safeguarding the intellectual property
of a pretrained model from malicious KD.
SDB pairs each wrapped model with a randomly-generated key,
issued to authorized users only,
and permits only authorized users
to conduct knowledge transfer from the model.
By contrast, an unauthorized KD attempt
would lead to a poorly-behaved student model.
Specifically, we deploy
three strategies in SDB to achieve our goal, namely
key embedding, knowledge disturbance, and knowledge preservation.
Experimental results over various datasets
and network architectures validate the 
effectiveness of SDB:
unauthorized KDs from the wrapped model yields
a significant performance drop,
while authorized KDs in fact preserve or enhance
the accuracy.
In our future work, we will explore deploying
SDB to edge terminals, and focus on its
applications over compact networks.

{\small
\begin{thebibliography}{35}
\providecommand{\natexlab}[1]{#1}

\bibitem[{Adi et~al.(2018)Adi, Baum, Ciss{\'e}, Pinkas, and
  Keshet}]{Adi2018TurningYW}
Adi, Y.; Baum, C.; Ciss{\'e}, M.; Pinkas, B.; and Keshet, J. 2018.
\newblock Turning Your Weakness Into a Strength: Watermarking Deep Neural
  Networks by Backdooring.
\newblock In \emph{USENIX Security Symposium}.

\bibitem[{Bagdasaryan et~al.(2020)Bagdasaryan, Veit, Hua, Estrin, and
  Shmatikov}]{Bagdasaryan2020HowTB}
Bagdasaryan, E.; Veit, A.; Hua, Y.; Estrin, D.; and Shmatikov, V. 2020.
\newblock How To Backdoor Federated Learning.
\newblock In \emph{AISTATS}.

\bibitem[{Barni, Kallas, and Tondi(2019)}]{Barni2019ANB}
Barni, M.; Kallas, K.; and Tondi, B. 2019.
\newblock A New Backdoor Attack in CNNS by Training Set Corruption Without
  Label Poisoning.
\newblock \emph{2019 IEEE International Conference on Image Processing (ICIP)},
  101--105.

\bibitem[{Chen et~al.(2020)Chen, Mei, Wang, Feng, and Chen}]{Chen2020OnlineKD}
Chen, D.; Mei, J.-P.; Wang, C.; Feng, Y.; and Chen, C. 2020.
\newblock Online Knowledge Distillation with Diverse Peers.
\newblock \emph{AAAI Conference on Artificial Intelligence}.

\bibitem[{Chen et~al.(2017)Chen, Choi, Yu, Han, and
  Chandraker}]{Chen2017LearningEO}
Chen, G.; Choi, W.; Yu, X.; Han, T.~X.; and Chandraker, M. 2017.
\newblock Learning Efficient Object Detection Models with Knowledge
  Distillation.
\newblock In \emph{Neural Information Processing Systems}.

\bibitem[{Dai, Chen, and Li(2019)}]{Dai2019ABA}
Dai, J.; Chen, C.; and Li, Y. 2019.
\newblock A Backdoor Attack Against LSTM-Based Text Classification Systems.
\newblock \emph{IEEE Access}, 7: 138872--138878.

\bibitem[{Fan et~al.(2021)Fan, Ng, Chan, and Yang}]{Fan2021DeepIPDN}
Fan, L.; Ng, K.~W.; Chan, C.~S.; and Yang, Q. 2021.
\newblock DeepIP: Deep Neural Network Intellectual Property Protection with
  Passports.
\newblock \emph{IEEE transactions on pattern analysis and machine
  intelligence}, PP.

\bibitem[{Gao et~al.(2017)Gao, Guo, Li, and Nevatia}]{gao2017knowledge}
Gao, J.; Guo, Z.; Li, Z.; and Nevatia, R. 2017.
\newblock Knowledge Concentration: Learning 100K Object Classifiers in a Single
  CNN.
\newblock \emph{arXiv}.

\bibitem[{Han et~al.(2020)Han, Song, Yao, Xu, and Nie}]{Han2020NeuralCM}
Han, X.; Song, X.; Yao, Y.; Xu, X.-S.; and Nie, L. 2020.
\newblock Neural Compatibility Modeling With Probabilistic Knowledge
  Distillation.
\newblock \emph{IEEE Transactions on Image Processing}, 29: 871--882.

\bibitem[{He et~al.(2016)He, Zhang, Ren, and Sun}]{He2016IdentityMI}
He, K.; Zhang, X.; Ren, S.; and Sun, J. 2016.
\newblock Identity Mappings in Deep Residual Networks.
\newblock \emph{ECCV}, abs/1603.05027.

\bibitem[{Hinton, Vinyals, and Dean(2015)}]{hinton2015distilling}
Hinton, G.~E.; Vinyals, O.; and Dean, J. 2015.
\newblock Distilling the Knowledge in a Neural Network.
\newblock \emph{Neural Information Processing Systems}.

\bibitem[{Huang et~al.(2018)Huang, You, Chen, Qian, and
  Yu}]{Huang2018KnowledgeDF}
Huang, M.; You, Y.; Chen, Z.; Qian, Y.; and Yu, K. 2018.
\newblock Knowledge Distillation for Sequence Model.
\newblock In \emph{INTERSPEECH}.

\bibitem[{Lange et~al.(2021)Lange, Aljundi, Masana, Parisot, Jia, Leonardis,
  Slabaugh, and Tuytelaars}]{Lange2021ACL}
Lange, M.~D.; Aljundi, R.; Masana, M.; Parisot, S.; Jia, X.; Leonardis, A.;
  Slabaugh, G.; and Tuytelaars, T. 2021.
\newblock A continual learning survey: Defying forgetting in classification
  tasks.
\newblock \emph{IEEE Transactions on Pattern Analysis and Machine
  Intelligence}, PP.

\bibitem[{Liao et~al.(2020)Liao, Zhong, Squicciarini, Zhu, and
  Miller}]{Liao2020BackdoorEI}
Liao, C.; Zhong, H.; Squicciarini, A.; Zhu, S.; and Miller, D.~J. 2020.
\newblock Backdoor Embedding in Convolutional Neural Network Models via
  Invisible Perturbation.
\newblock \emph{Proceedings of the Tenth ACM Conference on Data and Application
  Security and Privacy}.

\bibitem[{Liu et~al.(2020)Liu, Ma, Bailey, and Lu}]{Liu2020ReflectionBA}
Liu, Y.; Ma, X.; Bailey, J.; and Lu, F. 2020.
\newblock Reflection Backdoor: A Natural Backdoor Attack on Deep Neural
  Networks.
\newblock In \emph{European Conference on Computer Vision}.

\bibitem[{Ma et~al.(2021)Ma, Chen, Hu, You, Xie, and
  Wang}]{Ma2021UndistillableMA}
Ma, H.; Chen, T.; Hu, T.-K.; You, C.; Xie, X.; and Wang, Z. 2021.
\newblock Undistillable: Making A Nasty Teacher That CANNOT teach students.
\newblock \emph{International Conference on Learning Representations}.

\bibitem[{Ma et~al.(2018)Ma, Zhang, Zheng, and Sun}]{Ma2018ShuffleNetVP}
Ma, N.; Zhang, X.; Zheng, H.; and Sun, J. 2018.
\newblock ShuffleNet V2: Practical Guidelines for Efficient CNN Architecture
  Design.
\newblock \emph{ECCV}, abs/1807.11164.

\bibitem[{Merrer, P{\'e}rez, and Tr{\'e}dan(2019)}]{Merrer2019AdversarialFS}
Merrer, E.~L.; P{\'e}rez, P.; and Tr{\'e}dan, G. 2019.
\newblock Adversarial Frontier Stitching for Remote Neural Network
  Watermarking.
\newblock \emph{Neural Computing and Applications}, 32: 9233--9244.

\bibitem[{Mirzadeh et~al.(2020)Mirzadeh, Farajtabar, Li, Levine, Matsukawa, and
  Ghasemzadeh}]{Mirzadeh2020ImprovedKD}
Mirzadeh, S.; Farajtabar, M.; Li, A.; Levine, N.; Matsukawa, A.; and
  Ghasemzadeh, H. 2020.
\newblock Improved Knowledge Distillation via Teacher Assistant.
\newblock In \emph{AAAI Conference on Artificial Intelligence}.

\bibitem[{Nagai et~al.(2018)Nagai, Uchida, Sakazawa, and
  Satoh}]{Nagai2018DigitalWF}
Nagai, Y.; Uchida, Y.; Sakazawa, S.; and Satoh, S. 2018.
\newblock Digital watermarking for deep neural networks.
\newblock \emph{International Journal of Multimedia Information Retrieval}, 7:
  3--16.

\bibitem[{Park et~al.(2019)Park, Kim, Lu, and Cho}]{Park2019RelationalKD}
Park, W.; Kim, D.; Lu, Y.; and Cho, M. 2019.
\newblock Relational Knowledge Distillation.
\newblock \emph{Computer Vision and Pattern Recognition}, 3962--3971.

\bibitem[{Rosenfeld and Tsotsos(2020)}]{Rosenfeld2020IncrementalLT}
Rosenfeld, A.; and Tsotsos, J.~K. 2020.
\newblock Incremental Learning Through Deep Adaptation.
\newblock \emph{IEEE Transactions on Pattern Analysis and Machine
  Intelligence}, 42: 651--663.

\bibitem[{Saha, Subramanya, and Pirsiavash(2020)}]{Saha2020HiddenTB}
Saha, A.; Subramanya, A.; and Pirsiavash, H. 2020.
\newblock Hidden Trigger Backdoor Attacks.
\newblock In \emph{AAAI}.

\bibitem[{Sandler et~al.(2018)Sandler, Howard, Zhu, Zhmoginov, and
  Chen}]{Sandler2018MobileNetV2IR}
Sandler, M.; Howard, A.~G.; Zhu, M.; Zhmoginov, A.; and Chen, L.-C. 2018.
\newblock MobileNetV2: Inverted Residuals and Linear Bottlenecks.
\newblock \emph{2018 IEEE/CVF Conference on Computer Vision and Pattern
  Recognition}, 4510--4520.

\bibitem[{Shen et~al.(2019)Shen, Wang, Song, Sun, and
  Song}]{shen2019amalgamating}
Shen, C.; Wang, X.; Song, J.; Sun, L.; and Song, M. 2019.
\newblock Amalgamating knowledge towards comprehensive classification.
\newblock In \emph{AAAI Conference on Artificial Intelligence}, 3068--3075.

\bibitem[{Shibata et~al.(2021)Shibata, Irie, Ikami, and
  Mitsuzumi}]{Shibata2021LearningWS}
Shibata, T.; Irie, G.; Ikami, D.; and Mitsuzumi, Y. 2021.
\newblock Learning with Selective Forgetting.
\newblock In \emph{International Joint Conference on Artificial Intelligence}.

\bibitem[{Turner, Tsipras, and Madry(2018)}]{Turner2018CleanLabelBA}
Turner, A.; Tsipras, D.; and Madry, A. 2018.
\newblock Clean-Label Backdoor Attacks.

\bibitem[{Wang et~al.(2019)Wang, Yao, Shan, Li, Viswanath, Zheng, and
  Zhao}]{Wang2019NeuralCI}
Wang, B.; Yao, Y.; Shan, S.; Li, H.; Viswanath, B.; Zheng, H.; and Zhao, B.~Y.
  2019.
\newblock Neural Cleanse: Identifying and Mitigating Backdoor Attacks in Neural
  Networks.
\newblock \emph{IEEE Symposium on Security and Privacy}, 707--723.

\bibitem[{Xu et~al.(2018)Xu, Ouyang, Wang, and Sebe}]{Xu2018PADNetMG}
Xu, D.; Ouyang, W.; Wang, X.; and Sebe, N. 2018.
\newblock PAD-Net: Multi-tasks Guided Prediction-and-Distillation Network for
  Simultaneous Depth Estimation and Scene Parsing.
\newblock \emph{Computer Vision and pattern recognition}, 675--684.

\bibitem[{Yu et~al.(2017)Yu, Liu, Wang, and Tao}]{WangCVPR17}
Yu, X.; Liu, T.; Wang, X.; and Tao, D. 2017.
\newblock On Compressing Deep Models by Low Rank and Sparse Decomposition.
\newblock \emph{Computer Vision and pattern recognition}, 67--76.

\bibitem[{Zagoruyko and Komodakis(2017)}]{Zagoruyko2017AT}
Zagoruyko, S.; and Komodakis, N. 2017.
\newblock Paying More Attention to Attention: Improving the Performance of
  Convolutional Neural Networks via Attention Transfer.
\newblock In \emph{International Conference on Learning Representations}.

\bibitem[{Zhang et~al.(2018{\natexlab{a}})Zhang, Ciss{\'e}, Dauphin, and
  Lopez-Paz}]{Zhang2018mixupBE}
Zhang, H.; Ciss{\'e}, M.; Dauphin, Y.; and Lopez-Paz, D. 2018{\natexlab{a}}.
\newblock mixup: Beyond Empirical Risk Minimization.
\newblock \emph{International Conference on Learning Representations}.

\bibitem[{Zhang et~al.(2018{\natexlab{b}})Zhang, Gu, Jang, Wu, Stoecklin,
  Huang, and Molloy}]{Zhang2018ProtectingIP}
Zhang, J.; Gu, Z.; Jang, J.; Wu, H.; Stoecklin, M.; Huang, H.; and Molloy, I.
  2018{\natexlab{b}}.
\newblock Protecting Intellectual Property of Deep Neural Networks with
  Watermarking.
\newblock \emph{Proceedings of the 2018 on Asia Conference on Computer and
  Communications Security}.

\bibitem[{Zhang et~al.(2021)Zhang, Jia, Wang, and Gong}]{Zhang2021BackdoorAT}
Zhang, Z.; Jia, J.; Wang, B.; and Gong, N. 2021.
\newblock Backdoor Attacks to Graph Neural Networks.
\newblock \emph{Proceedings of the 26th ACM Symposium on Access Control Models
  and Technologies}.

\bibitem[{Zhao et~al.(2020)Zhao, Ma, Zheng, Bailey, Chen, and
  Jiang}]{Zhao2020CleanLabelBA}
Zhao, S.; Ma, X.; Zheng, X.; Bailey, J.; Chen, J.; and Jiang, Y. 2020.
\newblock Clean-Label Backdoor Attacks on Video Recognition Models.
\newblock \emph{Conference on Computer Vision and Pattern Recognition},
  14431--14440.

\end{thebibliography}

}
\end{document}